\documentclass[12pt]{article}
\usepackage{epsfig}
\newcommand{\be}{\begin{equation}}

\newcommand{\ee}{\end{equation}}
\newcommand{\ben}{\begin{eqnarray*}}
\newcommand{\een}{\end{eqnarray*}}
\newcommand{\stackeven}[2]{{{}_{\displaystyle{#1}}\atop\displaystyle{#2}}}

\newcommand{\gsim}{\stackeven{>}{\sim}}
\newcommand{\un}[1]{\underline{#1}}
\newcommand{\as}{\alpha_s}

\newcommand{\bas}{{\overline\alpha}_s}
\def\eq#1{{Eq.~(\ref{#1})}}
\def\fig#1{{Fig.~\ref{#1}}}

\begin{document}
\title{{\bf Inclusive Gluon Production In High Energy\vspace*{5mm} Onium-Onium Scattering 
\\[1.5cm] }}
\author{
{\bf Yuri V.\ Kovchegov\thanks{e-mail:
yuri@mps.ohio-state.edu}}~\\[1cm] {\it\small Department of Physics,
The Ohio State University}\\ {\it\small Columbus, OH 43210,
USA}\\[5mm]}

\date{August 2005}

\maketitle

\thispagestyle{empty}

\begin{abstract}
We calculate the inclusive single-gluon production cross section in
high energy onium-onium scattering including pomeron loop effects. The
resulting inclusive cross section is given by the $k_T$-factorization
formula with one of the unintegrated gluon distribution functions
depending on the total onium-onium scattering cross section, which
includes all pomeron loops and has to be found independently. We
discuss the limits of applicability of our result and argue that they
are given by the limits of applicability of pomeron loop resummation
approach. Since the obtained $k_T$-factorization formula is infrared
divergent we conclude that, in order to consistently calculate the
(infrared-finite) gluon production cross section in onium-onium
scattering, one has to include corrections going beyond the pomeron
loop approximation.
\end{abstract}

\thispagestyle{empty}

\newpage

\setcounter{page}{1}


\section{Introduction}

Recently there has been a lot of renewed interest in resummation of
the so-called ``pomeron loop'' diagrams for high energy scattering
\cite{loop1,loop2,loop3,loop4,loop5,loop6,loop7,loop8}. These diagrams are 
particularly important for the calculation of the cross section for
high energy scattering of two quark-onia. It has been argued that the
growth of the onium-onium total scattering cross section with energy
due to the BFKL evolution \cite{BFKL} is unitarized (at fixed impact
parameter) by the multiple hard pomeron exchanges realized by the
pomeron loop diagrams \cite{dip1,dip2,dip3,NP,Salam}. Such
unitarization is qualitatively different from the unitarization of
total scattering cross section in deep inelastic scattering (DIS) on a
nucleus \cite{glr,mv,yuri,bal,JKLW,FILM}: in the onium-onium
scattering case one considers a collision of two small and dilute
objects, whereas in the case of DIS one of the scatterers is a large
and dense nucleus.

The contribution of multiple pomeron exchanges to the total
onium-onium scattering cross section has been studied within the
framework of Mueller's dipole model \cite{dip1} both analytically by
Mueller and Patel \cite{dip2} and Navelet and Peschanski
\cite{NP} and numerically by Salam in \cite{Salam}. The approach of 
\cite{dip2,NP,Salam} has recently been reformulated in the language of the 
Color Glass Condensate by Iancu and Mueller in \cite{loop1}. The main
idea behind the dipole model approach to onium-onium scattering is
reviewed below in Sect. 2. When viewed in the center-of-mass frame,
each of the colliding onia develops a dipole wave function. The
dipoles in the light cone wave functions of the onia interact pairwise
by two-gluon exchanges \cite{dip2,loop1}. This process is equivalent
to resumming pomeron loop diagrams in the traditional Feynman diagram
language and is illustrated below in \fig{oo2}.

Our goal in this paper is to calculate the single gluon inclusive
production cross section in the onium-onium scattering described in
the framework of Mueller's dipole model \cite{dip2,loop1}. This is an
interesting observable one could study in $\gamma^* \gamma^*$
scattering at the future linear colliders. The calculation turns out
to be analogous to the case of gluon production in DIS considered in
\cite{KT} and is outlined in Sect. 3 below. Using the real-virtual 
cancellations of some of the final state interactions observed in
\cite{CM} we arrive at the expression for the inclusive cross 
section of the single gluon production given by \eq{incl}. As one can
also see from a different form of \eq{incl} given by \eq{kt}, the
resulting gluon production cross section adheres to AGK cutting rules
\cite{AGK} and is given by the $k_T$-factorization expression
\cite{lr}, similar to the results of \cite{KT,Braun1,Braun2} for DIS 
and to \cite{Mar}. $k_T$-factorization appears here because of the
fact that, due to the nature of the approximation of
\cite{dip2,loop1}, one of the onia wave functions (the one closer to
the produced gluon in rapidity) appears to be unsaturated, i.e., has
no saturation effects in it and is described simply by the linear BFKL
evolution. This makes the problem effectively identical to the case of
DIS or $pA$ collisions.

$k_T$-factorization formula for the gluon production cross section is
known to diverge as $\sim 1/k_T^2$ in the infrared \cite{CCFM} (here
$k_T$ is the transverse momentum of the produced gluon). This
divergence is believed to be curable if one includes saturation
effects in both of the onia wave functions, which would lead to a
break-down of the $k_T$-factorization expression but would yield an
infrared-finite production cross section (similar observation appears
to hold for gluon production in nucleus-nucleus collisions
\cite{yuriaa,KV,Lappi}). For the onium-onium scattering, saturation in
both of the onia wave functions can be included by resumming all
pomeron loop diagrams with the loops spanning less than a half of the
total rapidity interval. However, as we discuss in Sect. 4, when
rapidity is high enough for such ``small'' pomeron loops to become
important, other corrections to pomeron loops approach become equally
important as well. Some of these non-pomeron loop corrections have
been discussed in \cite{loop6} to explain the discrepancy between the
results of \cite{loop5} and \cite{loop2}. In the traditional Feynman
diagram language these corrections can be associated with higher order
terms in the impact factors, zero-rapidity size pomeron loops and/or
with iterations of NLO and NNLO BFKL kernels in the LO BFKL ladder.
Therefore, we arrive at a somewhat unsettling conclusion that in order
to obtain an infrared-finite gluon production cross section in
onium-onium scattering in a consistent way one may have to go beyond
the pomeron loop approximation to high energy scattering.


\section{Onium-Onium Scattering at Intermediately High Energy}

\begin{figure}[h]
\begin{center}
\epsfxsize=4.5cm
\leavevmode
\hbox{\epsffile{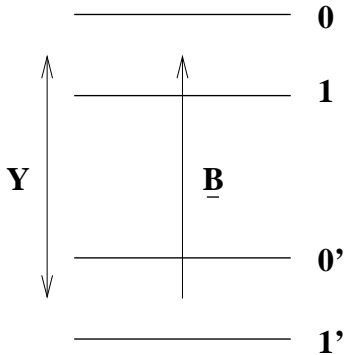}}
\end{center}
\caption{Onium-onium scattering as considered in the text.}
\label{oo1}
\end{figure}

Let us consider high energy scattering of two quarkonia in the
approximation put forward in \cite{dip2,dip3,loop1}. The notations we
will use are explained in \fig{oo1}. The onia have transverse sizes
${\un x}_{01} = {\un x}_0 - {\un x}_1$ and ${\un x}_{0'1'} = {\un
x}_{0'} - {\un x}_{1'}$ with ${\un x}_0 , {\un x}_1, {\un x}_{0'} ,
{\un x}_{1'}$ the transverse coordinates of quarks and anti-quarks in
both onia. The impact parameter of the collision is $\un B$ and the
rapidity interval between the onia is $Y$.

\begin{figure}[ht]
\begin{center}
\epsfig{file=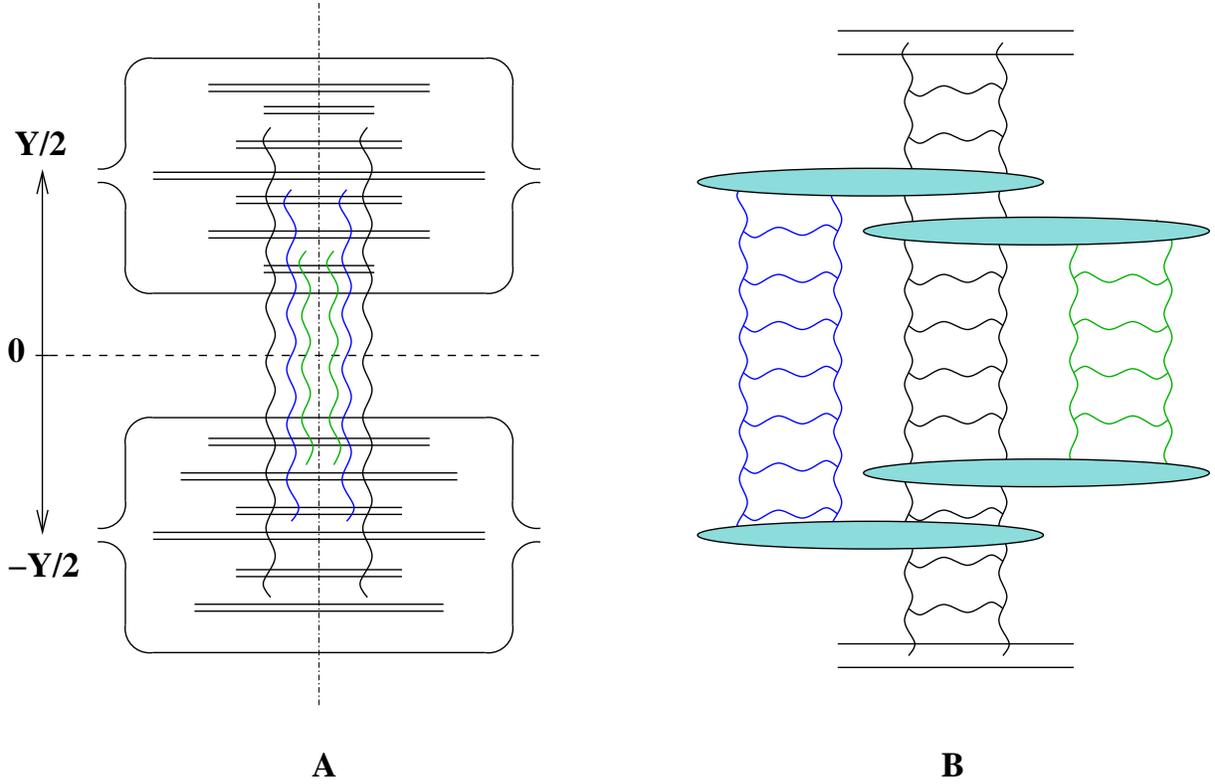,width=16cm}
\end{center}
\caption{Three-pomeron exchange diagram for onium-onium scattering in 
(A) dipole model and in (B) the Feynman diagram language. The double
lines in (A) denote gluons in the large-$N_c$ limit. The ovals in (B)
denote triple pomeron vertices. (Color on-line.)}
\label{oo2}
\end{figure}

The dipole model description of high energy onium-onium scattering
\cite{dip2,dip3,loop1} is illustrated in \fig{oo2}A. In the center 
of mass frame, each of the onia develops a system of color dipoles in
its light-cone wave function before the collisions. At the time of the
collision (denoted by dash-dotted line in \fig{oo2}A) the dipoles in
both onia interact with each other pairwise by two-gluon exchanges. In
\fig{oo2}A we show the case when three pairs of dipoles interact with
each other. This corresponds to three-pomeron exchange in the
traditional Feynman diagram language, which is illustrated in
\fig{oo2}B. In the dipole model \cite{dip2,dip3}, the BFKL evolution 
\cite{BFKL}, which is included in the ladder diagrams in \fig{oo2}B,
comes in through the dipole evolution in the light cone wave functions
of the onia, which leads to generation of the interacting dipoles. 

The dipole wave function of an onium is described by the generating
functional for dipoles $Z ({\un x}_{01}, {\un b}, Y, u)$, which obeys
the following evolution equation \cite{dip1,dip3}
\begin{eqnarray*}
  Z ({\un x}_{01}, {\un b}, Y, u) = u ({\un x}_{01}, {\un b}) 
  \, e^{- 2 \, \bas \, \ln \left( \frac{x_{01}}{\rho}
    \right) Y}
  + \frac{\bas}{2 \, \pi} \int_0^Y d y \, e^{- 2 \, \bas \, 
\ln \left( \frac{x_{01}}{\rho} \right) (Y - y)}
\end{eqnarray*}
\begin{eqnarray}\label{ZZ}
  \times \int_\rho d^2 x_2 \frac{x_{01}^2}{x_{02}^2 \, x_{12}^2}
  \, Z ({\un x}_{02}, {\un b} + \frac{1}{2}{\un x}_{12}, y, u) \, Z
  ({\un x}_{12}, {\un b} -\frac{1}{2}{\un x}_{20}, y, u)
\end{eqnarray}
with 
\be
\bas \equiv \frac{\as \, N_c}{\pi}
\ee
and $u ({\un x}_{01}, {\un b})$ a function describing a dipole of
transverse size ${\un x}_{01}$ at impact parameter $\un b$
\cite{dip3}.

Using $Z ({\un x}_{01}, {\un b}, Y, u)$ one can define the density of
$k$ dipoles in the onium wave function by \cite{dip2}
\be\label{nk}
n_k ({\un x}_{01}, {\un b}, Y; {\un r}_1, {\un b}_1; \ldots ; {\un
r}_k, {\un b}_k) \, = \,
\frac{1}{k!} \, \frac{\delta^k Z ({\un x}_{01}, {\un b}, Y, u)}{\delta u 
({\un r}_1, {\un b}_1) \, \ldots \, \delta u ({\un r}_k, {\un b}_k)}
\Bigg|_{u=1}.
\ee
If normalized appropriately, $n_k$ would have the meaning of {\sl
inclusive} probability of finding $k$ dipoles of given sizes ${\un
r}_i$'s at impact parameters ${\un b}_i$'s along with any number of
extra dipoles. Note that $n_k$, as defined by \eq{nk}, is normalized
to give density of $k$ dipoles rather than probability.

Let us denote the forward amplitude of dipole-dipole scattering
mediated by a two-gluon exchange as
\be\label{ddef}
- d ({\un r}, {\un r}', {\un b}) \, = \, \as^2 \, \frac{C_F}{N_c} \,
\ln^2 \left( \frac{|{\un b} + (1/2) {\un r} + (1/2) {\un r}' | \, 
|{\un b} - (1/2) {\un r} - (1/2) {\un r}'|}{|{\un b} + 
(1/2) {\un r} - (1/2) {\un r}' | \, 
|{\un b} - (1/2) {\un r} + (1/2) {\un r}'|} \right).
\ee
Here ${\un r}$ and ${\un r}'$ are transverse separations of the two
dipoles and $\un b$ is their impact parameter.

With the aid of the definition from \eq{ddef} we can write the forward
scattering amplitude for onium-onium scattering, which includes
small-$x$ evolution, as
\ben
- D ({\un x}_{01}, {\un x}_{0'1'}, {\un B}, Y) \, = \, -1 +
\sum_{k=1}^\infty \, k! \, \int d^2 r_1 d^2 b_1 \ldots d^2 r_k d^2 b_k \, 
d^2 r'_1 d^2 b'_1 \ldots d^2 r'_k d^2 b'_k 
\een
\ben
\times \, n_k ({\un x}_{01}, {\un
B}, Y/2; {\un r}_1, {\un b}_1; \ldots ; {\un r}_k, {\un b}_k) \, n_k
({\un x}_{0'1'}, {\un 0}, - Y/2; {\un r}'_1, {\un b}'_1; \ldots ; {\un
r}'_k, {\un b}'_k) 
\een
\be\label{D1}
\times \, (-1)^k \, d ({\un r}_1, {\un r}'_1, {\un b}_1 - {\un b}'_1) \, \ldots \,
d ({\un r}_k, {\un r}'_k, {\un b}_k - {\un b}'_k), 
\ee
where we chose the impact parameters of dipoles $01$ and $0'1'$ to be
$\un B$ and zero correspondingly. The notations of \eq{D1} are
explained in \fig{oo1}.

Using \eq{nk} in \eq{D1} we re-write it as 
\ben
- D ({\un x}_{01}, {\un x}_{0'1'}, {\un B}, Y) \, = \, -1 +
\sum_{k=1}^\infty \, \frac{1}{k!} \, \int d^2 r_1 d^2 b_1 \ldots d^2 r_k d^2 b_k \, 
d^2 r'_1 d^2 b'_1 \ldots d^2 r'_k d^2 b'_k 
\een
\ben
\times \, \frac{\delta^k Z ({\un x}_{01}, {\un B}, Y/2, u)}{\delta u 
({\un r}_1, {\un b}_1) \, \ldots \, \delta u ({\un r}_k, {\un b}_k)}
\Bigg|_{u=1} \, \frac{\delta^k Z ({\un x}_{0'1'}, {\un 0}, - Y/2, v)}{\delta v 
({\un r}'_1, {\un b}'_1) \, \ldots \, \delta v ({\un r}'_k, {\un b}'_k)}
\Bigg|_{v=1}
\een
\be\label{D2}
\times \, (-1)^k \, d ({\un r}_1, {\un r}'_1, {\un b}_1 - {\un b}'_1) \, \ldots \,
d ({\un r}_k, {\un r}'_k, {\un b}_k - {\un b}'_k),
\ee
where we relabeled functions $u$ by $v$ for the dipole $0'1'$. If one
could solve \eq{ZZ} for an arbitrary function $u$, then the solution
could be used in \eq{D2} to construct the forward amplitude of
onium-onium scattering. Unfortunately no general analytical solution
of \eq{ZZ} for an arbitrary function $u$ is known.

\eq{D2} can be recast in a more compact form:
\ben
D ({\un x}_{01}, {\un x}_{0'1'}, {\un B}, Y) \, = \, 1 - \Bigg\{ \exp 
\left[ - \int d^2 r \, d^2 b  \, d^2 r' \, d^2 b' \ d ({\un r}, {\un r}', 
{\un b} - {\un b}') \, \frac{\delta^2}{\delta u ({\un r}, {\un b}) \,
\delta v ({\un r}', {\un b}') } \right] 
\een
\be\label{D3}
\times \, Z ({\un x}_{01}, {\un B}, Y/2, u) 
\, Z ({\un x}_{0'1'}, {\un 0}, - Y/2, v) \Bigg\}\Bigg|_{u=1, \, v=1}.
\ee
This is our final expression for the onium-onium forward scattering
amplitude here. One could also try to rewrite \eq{D3} in terms of a
functional integral over functions $u$ and $v$: however, such integral
representation appears to require non-Gaussian weight functional,
which is probably related to the non-Gaussian noise discussed in
\cite{loop2} within the context of adding pomeron loop corrections to JIMWLK 
evolution equations \cite{JKLW,FILM}.

The question of whether $D ({\un x}_{01}, {\un x}_{0'1'}, {\un B}, Y)$
gives a unitary total onium-onium scattering cross section has been
addressed numerically by Salam in \cite{Salam}. It was demonstrated by
Monte Carlo simulations in \cite{Salam} that $D ({\un x}_{01}, {\un
x}_{0'1'}, {\un B}, Y)$ is unitary at a fixed impact parameter $\un
b$. 
To simplify the analysis of $D ({\un x}_{01}, {\un x}_{0'1'}, {\un B},
Y)$ in \eq{D3} it would be very useful to find an
integral/differential equation governing the high energy evolution of
$D ({\un x}_{01}, {\un x}_{0'1'}, {\un B}, Y)$, similar to the case of
DIS considered in \cite{yuri,bal}. However, despite many attempts, no
such equation for $D ({\un x}_{01}, {\un x}_{0'1'}, {\un B}, Y)$ has
been found at this time.

\begin{figure}[ht]
\begin{center}
\epsfxsize=9.5cm
\leavevmode
\hbox{\epsffile{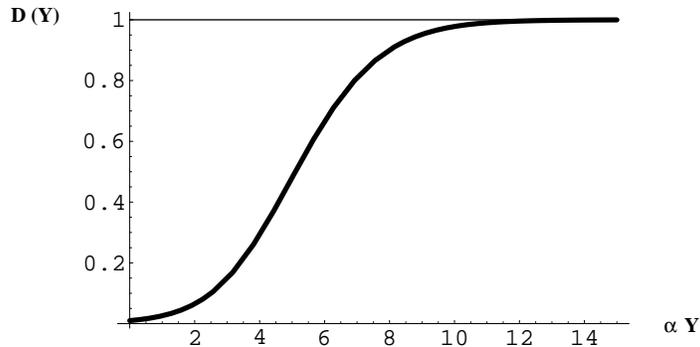}}
\end{center}
\caption{Onium-onium scattering amplitude $D(Y)$ from the toy model result 
in \eq{Dtoy} plotted as a function of $\as \, Y$ for $d=0.01$ (thick
line). The thin horizontal line denotes the black disk limit.}
\label{dunit}
\end{figure}

We would illustrate the onset of unitarity in $D$ from \eq{D3} by
using a simple toy model, originally proposed by Mueller in
\cite{dip3}. Suppressing the transverse coordinate dependence we write
the toy model analogue of \eq{ZZ} as \cite{dip3}
\be\label{ztoy}
\frac{d Z}{d Y} \, = \, \as \, Z^2 - \as Z
\ee
with the initial condition given by 
\be\label{toyic}
Z (Y=0, u) \, = \, u.
\ee
The solution of the toy evolution equation (\ref{ztoy}) satisfying the
initial condition (\ref{toyic}) is \cite{dip3}
\be\label{zsol}
Z (Y,u) \, = \, \frac{u}{u + (1-u) \, e^{\as \, Y}}.
\ee
The toy model analogue of \eq{D3} is
\be\label{dtoy}
D (Y) \, = \, 1 - \left\{ \exp \left[ - d \, \frac{\delta^2}{\delta u \,
\delta v} \right] \, Z (Y/2,u) \, Z (Y/2,v) \right\}\Bigg|_{u=1, \, v=1},
\ee
where $d$ is some {\sl positive} small number, $0 < d < 1$, and we put
$Y/2$ instead of $-Y/2$ in the argument of $Z (Y/2,v)$ to underline
the fact that it is the absolute value of the rapidity interval, and
not the rapidity itself, which matters there. Substituting $Z$ from
\eq{zsol} into \eq{dtoy}, expanding the exponential and performing the
differentiations with respect to $u$ and $v$ yields
\be\label{dser}
D (Y) \, = \, - \sum_{n=1}^\infty \, n! \, (-1)^n \, d^n \, e^{\as \,
Y} \, (1 - e^{\as \, Y /2})^{2 \, n - 2}.
\ee
The series in \eq{dser} is divergent: however, since $d > 0$, it is
Borel-summable. At large rapidities we can approximate $1 - e^{\as \,
Y /2} \approx - e^{\as \, Y /2}$ rewriting \eq{dser} as
\be
D (Y) \, \approx \, - \sum_{n=1}^\infty \, n! \, (-1)^n \, d^n \,
e^{n \, \as \, Y }
\ee
which after Borel resummation yields
\be\label{Dtoy}
D (Y) \, \approx \, 1 - \frac{1}{d \, e^{\as \, Y }} \, 
\exp \left( \frac{1}{d \, e^{\as
\, Y }} \right) \, \Gamma \left( 0, \frac{1}{d \, e^{\as \, Y }} \right).
\ee
The amplitude $D(Y)$ from \eq{Dtoy} is shown in \fig{dunit} as a
function of $\as \, Y$. One can see that at very large $Y$ the black
disk limit ($D=1$) is reached and is never exceeded, i.e., the
amplitude is unitary.

It is interesting to note that if one does not make the $1 - e^{\as \,
Y /2} \approx - e^{\as \, Y /2}$ approximation and Borel-resums the
full series in \eq{dser}, the resulting amplitude would violate
unitarity: as one increases rapidity it would grow, becoming greater
than $1$ at some large rapidity $Y$, and, after reaching a maximum it
would slowly decrease, approaching $1$ from above at asymptotically
high rapidities. Indeed the pomeron loop approximation of
\cite{dip2,loop1} consistently resums only the leading terms at high
energy, and has no control over the subleading terms, such as $1$ that
we neglected compared to $e^{\as \, Y /2}$. Therefore, such
pathological behavior of $D(Y)$ is probably due to the fact that the
toy model at hand, along with the approximation of \cite{dip2,loop1},
do not consistently resum the subleading terms. However, it may also
indicate the importance of subleading corrections for onium-onium
scattering at very high rapidities. After all, the same toy model
would always give a unitary cross section for DIS, where the
dipole-nucleus forward scattering amplitude is given by $N = 1-Z$
\cite{yuri,bal}.  We will return to this discussion below in Sect. 4.


\section{Inclusive Gluon Production}

\begin{figure}[h]
\begin{center}
\epsfxsize=6cm
\leavevmode
\hbox{\epsffile{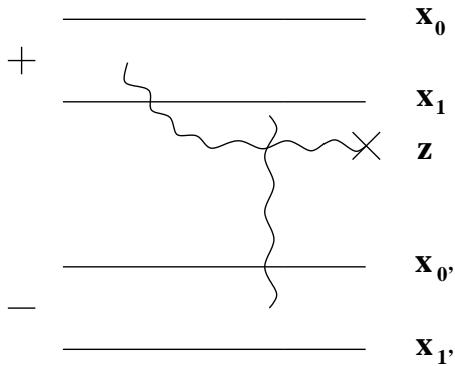}}
\end{center}
\caption{Lowest order gluon production in onium-onium scattering.}
\label{lo}
\end{figure}

To calculate single gluon production cross section in the
approximation to onium-onium scattering developed in \cite{dip2,loop1}
we start with the lowest order contribution. One of the corresponding
diagrams is shown in \fig{lo}. The produced gluon is denoted by the
cross. The disconnected $s$-channel (horizontal) gluon line in
\fig{lo} indicates that the gluon can be emitted off either the quark
or the anti-quark lines in the onium $01$. Disconnected $t$-channel
(vertical) gluon line denotes the sum of the gluon interactions with
the emitted gluon, quark and anti-quark in dipole $01$ at one end, and
with the quark and the anti-quark in dipole $0'1'$ at the other end.

The dipole model approach to onium-onium scattering \cite{dip2} was
developed in the Coulomb gauge, which is equivalent to the light cone
$A_+ =0$ gauge for the wave function of the onium moving in the light
cone ``$+$'' direction and to $ A_- =0$ gauge for the wave function of
the onium moving in the light cone ``$-$'' direction. In analyzing the
lowest order diagram from \fig{lo} we will work in the $A_+ =0$ light
cone gauge, where the $01$ onium is taken to be going in the ``$+$''
direction and the $0'1'$ onium is taken to be going in the ``$-$''
direction.

The lowest order gluon production cross section is well-known
\cite{KMW,KR,GM}. Following \cite{KM,KT} we can write it as (for a 
review of similar derivations see \cite{JMKrev})
\begin{eqnarray}\label{loincl}
\frac{d \sigma_{LO} ({\un x}_{01}, {\un x}_{0'1'})}{d^2 k \, dy \, d^2 B} 
\, = \,  \frac{\as C_F}{\pi^2} \, \frac{1}{(2 \pi)^2} \, \int \, 
d^2 z_1 \, d^2 z_2 \, e^{- i {\underline k} \cdot ({\underline z}_1 -
{\underline z}_2)} \, \sum_{i,j=0}^1 (-1)^{i+j}
\frac{{\underline z}_1- {\underline x}_i}{|{\underline z}_1-
{\underline x}_i|^2} \cdot
\frac{{\underline z}_2- {\underline x}_j}{|{\underline z}_2- {\underline
x}_j|^2} \nonumber \\
\times \bigg[ d_{Gq} \left({\underline z}_1 - {\underline x}_j, {\un x}_{0'1'},
\frac{1}{2} ({\underline z}_1 + {\underline x}_j) \right) + d_{Gq}
\left({\underline z}_2 - {\underline x}_i,  {\un x}_{0'1'}, \frac{1}{2} ({\underline
z}_2 + {\underline x}_i) \right) \nonumber \\ - d_{Gq}
\left({\underline z}_1 - {\underline z}_2,  {\un x}_{0'1'}, \frac{1}{2} ({\underline
z}_1 + {\underline z}_2) \right)  - d_{Gq} \left({\underline
x}_i - {\underline x}_j, {\un x}_{0'1'}, \frac{1}{2} ({\underline x}_i
+ {\underline x}_j) \right)\bigg],
\end{eqnarray}
where
\be
d_{Gq} ({\un r}, {\un r}', {\un b}) \, = \, \frac{N_c}{C_F} \, d ({\un
r}, {\un r}', {\un b}) \, = \, \as^2 \,
\ln^2 \left( \frac{|{\un b} + (1/2) {\un r} + (1/2) {\un r}' | \, 
|{\un b} - (1/2) {\un r} - (1/2) {\un r}'|}{|{\un b} + 
(1/2) {\un r} - (1/2) {\un r}' | \, 
|{\un b} - (1/2) {\un r} + (1/2) {\un r}'|} \right)
\ee
is the forward amplitude for scattering of an adjoint (gluon) dipole
of size $\un r$ on a fundamental (quark) dipole of size $\un r'$
separated by the impact parameter $\un b$ and mediated by a two-gluon
exchange. Again we assume that the impact parameter of the onium
$0'1'$ is zero. In \eq{loincl}, $\un k$ and $y$ are transverse
momentum and rapidity of the produced gluon, while ${\un z}_1$ and
${\un z}_2$ are its transverse coordinates, which are different on
both sides of the cut \cite{KM,KT,yuriaa,JMKrev}.

The task now is to include the effects of small-$x$ evolution taking
place in both of the onia's wave functions into \eq{loincl}. In our
approach we will be closely following the case of gluon production in
DIS considered previously in \cite{KT}. Just like in \fig{oo2} we will
consider onium-onium scattering in the center-of-mass frame, and, like
in \cite{dip2} we will be working in Coulomb gauge. We will consider
the case in which the produced gluon has rapidity $y$ such that $0 < y
< Y/2$, i.e., the gluon is emitted closer to onium $01$ than to $0'1'$
in the rapidity interval. The rapidity-dependent gluon emissions of
\fig{oo2} generating small-$x$ evolution can be divided into three 
categories: (i) emission of ``harder'' gluons, with rapidity $\tilde
y$ such that $y < {\tilde y} < Y/2$; (ii) emission of ``softer''
gluons in the same onium's ($01$) wave function, i.e., emissions in
the rapidity interval $0 < {\tilde y} < y$; (iii) emissions in the
wave function of onium $0'1'$, in the rapidity interval $-Y/2 <
{\tilde y} < 0$.

The analysis of category (i) and (ii) emissions is identical to that
for gluon production in DIS carried out in Sect. IIIA and IIIB of
\cite{KT} correspondingly and heavily relies on the real-virtual 
cancellations of the final state interactions discussed in
\cite{CM}. In \cite{KT} the onium develops a dipole wave function 
which later interacts with the target nucleus. If the onium is moving
along the light cone ``$+$'' direction, the interval of the light cone
time $x_+$ over which it interacts with the nucleus is negligibly
short compared to the time it takes for the dipole wave function to
develop \cite{dip1,KM,JMKrev}. This is very similar to onium-onium
scattering: the light cone interaction time between the two onia's
wave functions is negligible compared to the time required to develop
each of these wave functions. Thus, from the standpoint of one of
these wave functions (the one in onium $01$), interaction is
instantaneous, happening at, say, light cone time $x_+ =0$. Then,
emissions of harder gluons [type (i)] may happen either before or
after the interaction, at $x_+ <0$ and $x_+ > 0$ correspondingly. They
may be real (the gluon is present in $x_+ = + \infty$ final state) or
virtual (the gluon is absent in the final state), given by Figs. 2 and
4 in \cite{KT} correspondingly. Due to real-virtual cancellations of
\cite{CM}, the final state emissions/absorptions of such gluons
cancel, leaving only initial state radiation. The overall effect of
type (i) emissions is to develop a dipole cascade before the collision
which would lead to creation of the ``last'' dipole in which the
measured gluon is produced. In other words, instead of being emitted
by the quark and anti-quark in the original onium $01$, as shown in
\fig{lo} leading to \eq{loincl}, the measured gluon would not be emitted 
in some dipole $23$, generated by the dipole evolution.

The dipole number density at rapidity $y$ in the onium $01$ is
described by the quantity $n_1 ({\un x}_{01}, {\un B}, Y/2 -y; {\un
x}_{23}, (1/2) ({\un x}_2 + {\un x}_3))$ from \eq{nk} with
$k=1$. Using \eq{ZZ} we see that it obeys the following differential
equation \cite{dip2,dip3}, which is actually equivalent to the BFKL
equation \cite{BFKL}:
\ben
\frac{\partial n_1 ({\un x}_{01}, {\un B}, y; 
{\un x}_{23}, {\un b})}{\partial y} \, = \,
\frac{\as \, N_c}{2 \, \pi^2} \, 
\int d^2 x_2 \, \frac{x_{01}^2}{x_{20}^2 \, x_{21}^2} \, 
\bigg[ n_1 ({\un x}_{02}, {\un B} + \frac{1}{2} \, {\un x}_{21}, y;  
{\un x}_{23}, {\un b})
\een
\be\label{eqn}
+ n_1 ({\un x}_{12}, {\un B} + \frac{1}{2} \, {\un x}_{20}, y; {\un
x}_{23}, {\un b}) - n_1 ({\un x}_{01}, {\un B}, y; {\un x}_{23}, {\un
b}) \bigg]
\ee
with the initial condition 
\be
n_1 ({\un x}_{01}, {\un B}, y=0; {\un x}_{23}, {\un b}) \, = \, \delta^2 ({\un
x}_{01} - {\un x}_{23}) \, \delta^2 ({\un B} - {\un b}).
\ee
To include the effects of type (i) emissions one has to modify the
cross section in \eq{loincl} by \cite{KT,JMKrev}
\be\label{typei}
\frac{d \sigma_{LO} ({\un x}_{01}, {\un x}_{0'1'})}{d^2 k \, dy \, d^2 B} 
\, \rightarrow \, \int \, d^2 x_{23} \, d^2 b \, n_1 ({\un x}_{01}, {\un B}, 
Y/2 -y; {\un x}_{23}, {\un b} ) \, 
\frac{d \sigma_{LO} ({\un x}_{23}, {\un x}_{0'1'})}{d^2 k \, dy \, d^2 b},
\ee
where we are using the notations of \cite{JMKrev}, which are slightly
different from those of \cite{KT}.

The emissions of softer [type (ii)] gluons can again be analyzed
similar to the DIS case of \cite{KT}. Analogous to how it was done in
Sect. IIIB of \cite{KT}, one can show that the effect of type (ii)
emissions is to convert $d_{Gq}$ in \eq{loincl} into the gluon
dipole--quark dipole ($0'1'$) forward scattering amplitude with a
fully evolved wave function of the (upper) gluon dipole. However, to
complete the picture one has to understand what happens to the wave
function of the (lower) onium $0'1'$, i.e., one has to analyze type
(iii) emissions. This is rather straightforward: the lower onium
$0'1'$ develops a dipole wave function before the collision. In its
light cone time $x_-$ the collision is, again, instantaneous,
happening at $x_- =0$, after which it can again emit more
gluons. However, none of the gluons emitted in this lower onium wave
function is tagged upon, i.e., none of them is required to be present
in the final state. Therefore, the rules for developing the lower
onium wave functions are the same as for the calculation of total
onium-onium scattering cross sections: all final state emissions
cancel \cite{CM}, leaving only the initial state emissions shown in
\fig{oo2}A and described by the generating functional $Z$ from
\eq{ZZ}. [We note also that there is no ``cross-talk'' ($s$-channel 
mergers) between emissions in the upper and lower onia in the leading
logarithmic approximation considered here.] Combining what we know now
about type (ii) and type (iii) gluon emissions we conclude that they
modify $d_{Gq}$ in \eq{loincl} into the gluon dipole--quark dipole
forward scattering amplitude with a fully evolved wave functions of
both gluon and quark dipoles. To include type (ii) and type (iii)
gluon emissions into \eq{loincl} we thus have to replace
\be\label{dgrepl}
d_{Gq} ({\un r}, {\un r}', {\un b}) \, \rightarrow \, D_{Gq} \left( {\un r},
{\un r}', {\un b}, y + \frac{Y}{2} \right)
\ee
in it. The quantity $D_{Gq}$ can be defined by analogy with $D$ from
\eq{D3} by noting that, in the large-$N_c$ limit, the gluon dipole is 
equivalent to two quark dipoles, such that the corresponding
generating functional is just a square of the generating functional
for the quark dipole, 
\be
Z_G ({\un x}_{01}, {\un B}, Y/2, u) \, = \, [Z ({\un x}_{01}, {\un B}, Y/2, u)]^2,
\ee
so that
\ben
D_{Gq} ({\un x}_{01}, {\un x}_{0'1'}, {\un B}, Y) \, = \, 1 - \Bigg\{ \exp 
\left[ - \int d^2 r \, d^2 b  \, d^2 r' \, d^2 b' \ d ({\un r}, {\un r}', 
{\un b} - {\un b}') \, \frac{\delta^2}{\delta u ({\un r}, {\un b}) \,
\delta v ({\un r}', {\un b}') } \right] 
\een
\be\label{DGq}
\times \, Z ({\un x}_{01}, {\un B}, Y/2, u) \, Z ({\un x}_{01}, {\un B}, Y/2, u) 
\, Z ({\un x}_{0'1'}, {\un 0}, - Y/2, v) \Bigg\}\Bigg|_{u=1, \, v=1}.
\ee

Combining Eqs. (\ref{typei}) and (\ref{dgrepl}) we finally write for
the inclusive gluon production cross section in onium-onium collisions
\begin{eqnarray}\label{incl}
\frac{d \sigma ({\un x}_{01}, {\un x}_{0'1'})}{d^2 k \, dy \, d^2 B} 
\, = \,  \int \, d^2 x_{2} \, d^2 x_3 \ n_1 \left( {\un x}_{01}, {\un B}, 
\frac{Y}{2} -y; {\un x}_{23}, \frac{1}{2} ({\un x}_2 + {\un x}_3) \right) \nonumber \\ 
\times \, \frac{\as C_F}{\pi^2} \, \frac{1}{(2 \pi)^2} \, 
\int \, d^2 z_1 \, d^2 z_2 \, e^{- i {\underline k} \cdot ({\underline z}_1 -
{\underline z}_2)} \, \sum_{i,j=2}^3 (-1)^{i+j}
\frac{{\underline z}_1- {\underline x}_i}{|{\underline z}_1-
{\underline x}_i|^2} \cdot
\frac{{\underline z}_2- {\underline x}_j}{|{\underline z}_2- {\underline
x}_j|^2} \nonumber \\
\times \bigg[ D_{Gq} \left({\underline z}_1 - {\underline x}_j, {\un x}_{0'1'},
\frac{1}{2} ({\underline z}_1 + {\underline x}_j) , y + \frac{Y}{2} \right) + D_{Gq}
\left({\underline z}_2 - {\underline x}_i,  {\un x}_{0'1'}, \frac{1}{2} ({\underline
z}_2 + {\underline x}_i), y + \frac{Y}{2} \right) \nonumber \\ -
D_{Gq} \left({\underline z}_1 - {\underline z}_2, {\un x}_{0'1'},
\frac{1}{2} ({\underline z}_1 + {\underline z}_2), y + \frac{Y}{2}
\right) - D_{Gq} \left({\underline x}_i - {\underline x}_j, {\un
x}_{0'1'}, \frac{1}{2} ({\underline x}_i + {\underline x}_j), y +
\frac{Y}{2} \right)\bigg]. 
\end{eqnarray}
This is our main result. The above formula is valid for $0 < y < Y/2$,
and can be easily modified to give the production cross section for
$-Y/2 < y <0$ by switching
\be
y \, \leftrightarrow \, - y, \hspace*{1cm} {\un x}_{01}
\leftrightarrow {\un x}_{0'1'}.
\ee
If $y=0$ (gluon is produced at mid-rapidity) \eq{incl} becomes
symmetric, since $D_{gq}$ should contain only the linear BFKL
evolution, due to the fact that the approximation to onium-onium
scattering employed here \cite{dip2,loop1} does not consistently resum
pomeron loops with rapidity extent less or equal to $Y/2$ (see
\fig{ploop1} below). While formally \eq{incl} with $y=0$ sums up some
pomeron loops of rapidity size $Y/2$ and smaller in $D_{Gq}$, as given
by \eq{DGq}, such loops were not consistently summed over in arriving at
\eq{incl}. Therefore, all but the linear BFKL pomeron terms in
$D_{Gq}$ should be neglected in \eq{incl} at $y=0$.

It is interesting to notice that the evolution between the produced
gluon and the closest onium (onium $01$ in \eq{incl}) is given by the
linear BFKL equation. This result is in accordance with the AGK
cutting rules \cite{AGK}. This is indeed surprising, particularly due
to the fact that AGK violations have been found recently in inclusive
production of $q \bar q$ pairs \cite{BGV} and in inclusive two-gluon
production \cite{JMK} in DIS and $pA$ collisions (see \cite{Braun2}
for a relevant discussion). However, the reason our result
(\ref{incl}) adheres to AGK rules is the same as for the single
inclusive gluon production cross section in DIS: due to the nature of
the approximation to the onium-onium scattering which we employed here
following \cite{dip2,loop1}, there is no difference between the wave
function of the onium closest to the produced gluon in rapidity and
the dipole wave function of the $q \bar q$ pair in DIS, considered in
\cite{KT}. Both wave functions only include pomeron splittings. This 
leads to the similarity between \eq{incl} and Eq. (30) in \cite{KT}
and to the fact that both cross sections adhere to AGK cutting
rules. While we can not explain why some inclusive cross sections
\cite{KT} adhere to AGK rules and some other cross sections
\cite{BGV,JMK} violate them, it is clear that the reason AGK rules
work for onium-onium scattering as well as for DIS is due to {\sl
moderately} high energy approximations applied with the non-linear
saturation effects being included only in one of the scatterers wave
functions.

Similar to the result of \cite{KT}, \eq{incl} can be recast in
$k_T$-factorization form \cite{lr}. First we note that, as long as
impact parameters are concerned, $n_1 ({\un x}, {\un B}, Y; {\un r},
{\un b})$ depends only on ${\un B} - \un b$. Therefore, 
\be
\int d^2 B \, n_1 \left( {\un x}_{01}, {\un B}, 
\frac{Y}{2} -y; {\un x}_{23}, \frac{1}{2} ({\un x}_2 + {\un x}_3) \right) 
\, \equiv \, n_1 \left( {\un x}_{01}, 
\frac{Y}{2} -y; {\un x}_{23} \right)
\ee
in \eq{incl} depends on ${\un x}_{23} \, = \, {\un x}_2 - {\un x}_3$,
but does not depend on ${\un x}_2 + {\un x}_3$. Using this fact we can
recast \eq{incl} into the following form, similar to how it was done
for Eq. (30) in \cite{KT} (note that now we do not assume that the
amplitude $D_{Gq}$ is independent of impact parameter, as we did in
\cite{KT}):
\ben
\frac{d \sigma ({\un x}_{01}, {\un x}_{0'1'})}{d^2 k \, dy} 
\, = \,   \frac{\as C_F}{(2 \, \pi)^3} \, \frac{1}{{\un k}^2} \, \int 
\, d^2 z \, \left[ \nabla_z^2 \int d^2 b \, 
D_{Gq} \left({\underline z}, {\un x}_{0'1'},
{\un b}, y + \frac{Y}{2} \right)\right] \, e^{- i {\un k} \cdot {\un
z}} \,
\een
\be\label{s1}
\times \, \left[ \nabla_z^2  \, \sum_{i,j=2}^3 (-1)^{i+j} \, \int d^2 x_{23} \, 
({\un z} - {\un x}_{ij})^2 \, \ln \left(
\frac{1}{|{\un z} - {\un x}_{ij}| \, \mu}\right) \, n_1 \left( {\un x}_{01}, 
\frac{Y}{2} -y; {\un x}_{23} \right) \right].
\ee
For derivation of \eq{s1} see Appendix A. Defining the unintegrated
gluon distribution functions in the onia by
\ben
\phi_{01} ({\un k}, y) \, = \, \frac{C_F}{\as \, (2 \pi)^3} \, \int d^2 z \, 
e^{- i {\un k} \cdot {\un z}} \, \nabla_z^2  \, 
\bigg[ \pi \, \as^2 \, \sum_{i,j=2}^3 (-1)^{i+j} \, \int d^2 x_{23} \, 
({\un z} - {\un x}_{ij})^2 \, \ln \left(
\frac{1}{|{\un z} - {\un x}_{ij}| \, \mu}\right) 
\een
\be
\times \, n_1 \left( {\un x}_{01},  y; {\un x}_{23} \right) \bigg]
\ee
and
\be
\phi_{0'1'} ({\un k}, y) \, = \, \frac{C_F}{\as \, (2 \pi)^3} \, 
\int d^2 z \, d^2 b \, e^{- i {\un k} \cdot {\un z}} \, \nabla_z^2 \, D_{Gq}
\left({\underline z}, {\un x}_{0'1'}, {\un b}, y \right)
\ee
we can rewrite \eq{s1} in the familiar $k_T$-factorized form
\cite{lr,Braun1,KT}
\be\label{kt}
\frac{d \sigma ({\un x}_{01}, {\un x}_{0'1'})}{d^2 k \, dy} 
\, = \,  \frac{2 \, \as}{C_F} \, \frac{1}{{\un k}^2} \, \int d^2 q \ 
\phi_{01} \left({\un q}, \frac{Y}{2} -y \right) \, \phi_{0'1'} 
\left({\un k} - {\un q}, y + \frac{Y}{2} \right).
\ee
We have thus shown that the inclusive gluon production cross section
in onium-onium scattering analyzed in the approximation of
\cite{dip2,loop1} is given by the $k_T$-factorization formula (\ref{kt}).


\section{Limits of Applicability of the Pomeron Loop Resummation}

\eq{kt} presents a problem. As long as $\phi_{01}$ and $\phi_{0'1'}$ 
are positive definite, the inclusive cross section in \eq{kt} diverges
as $\sim 1/{\un k}^2$ in the infrared (at low $k_T = |{\un k}|$). This
is a well-known problem of perturbative gluon production
\cite{KMW,KR,GM} resulting in the infinite total number of gluons produced. 
The divergence is usually $\sim 1/k_T^4$ at the lowest order (for $pp$
collisions) \cite{KMW,KR,GM} shown in \fig{lo}. It reduces to $\sim
1/k_T^2$ for $pA$ collisions and DIS \cite{KM}, when the saturation
effects are included in the target nucleus. There is a strong belief
in the community that including saturation effects both in the target
and in the projectile wave functions would eliminate the power-law
divergence completely. (Elimination of divergence would indeed require
the $k_T$-factorization formula to break down, since no matter what
the gluon distributions are it would always lead to $\sim 1/k_T^2$
singularity in the IR.) The belief is confirmed by numerical
\cite{KV,Lappi} and analytic \cite{yuriaa} analyses of gluon
production in $AA$ collisions. Therefore, one would expect that a {\sl
complete} description of onium-onium scattering should also lead to
infrared-regular gluon production cross section. The fact that our
result (\ref{kt}) does not give such cross section, indicates that the
approach to onium-onium collisions from \cite{dip2,loop1} applied here
is incomplete. (Indeed the same problem is present in DIS and $pA$
collisions in the approach used in \cite{yuri,KM,KT}.)

To see what is missing in the approach of \cite{dip2,loop1} let us
perform a few parametric estimates. First of all, a single BFKL
pomeron exchange over rapidity interval $Y$ is, parametrically, of the
order of
\be\label{pom}
\as^2 \, e^{(\alpha_P - 1) \, Y},
\ee
where
\be
\alpha_P - 1 \, = \, \frac{4 \, \as \, N_c}{\pi} \, \ln 2
\ee
is the intercept of the BFKL pomeron \cite{BFKL}. The exponent in
\eq{pom} comes from ladder rungs resummation with the factor of 
$\as^2$ in the traditional Feynman diagram language coming from
coupling of the ladder to the onia (impact factors). In the dipole
model the factor $\as^2$ comes from the two-gluon exchange interaction
between the two dipoles at the ends of dipole evolutions in both onia
[\eq{ddef}].

Pomeron loops become important in onium-onium scattering at rapidity
$Y_U$ at which the correction (\ref{pom}) brought in by an extra
ladder becomes of order one, i.e., when
\be\label{yudef}
\as^2 \, e^{(\alpha_P - 1) \, Y_U} \, \sim \, 1,
\ee
leading to \cite{dip3,yuri}
\be
Y_U \, \sim \, \frac{1}{\alpha_P - 1} \, \ln \frac{1}{\as^2}.
\ee
Therefore, for rapidities $Y \gsim Y_U$ pomeron loops become important
and have to be resummed.
\begin{figure}
\begin{center}
\epsfxsize=12cm
\leavevmode
\hbox{\epsffile{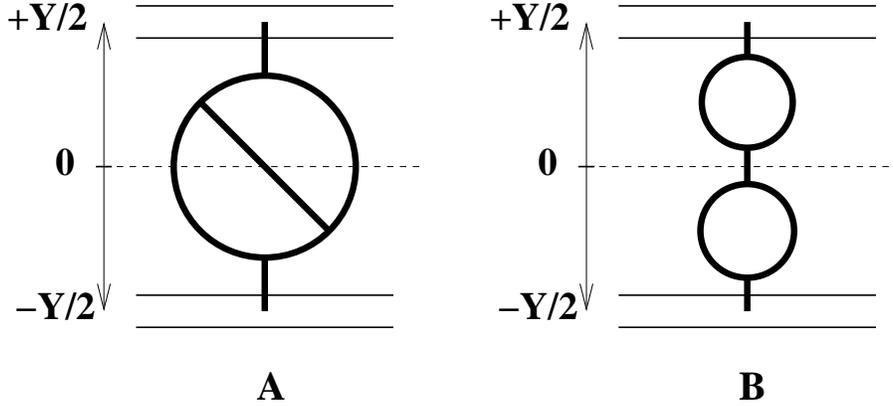}}
\end{center}
\caption{The pomeron loop diagrams included (A) and not included (B) in 
the approach to onium-onium scattering of \cite{dip2,loop1}. Thick
lines denote BFKL ladders.}
\label{ploop1}
\end{figure}

The types of pomeron loops included in the approach proposed in
\cite{dip2,loop1} are shown in \fig{ploop1}A. There the thick lines 
denote BFKL ladders (hard pomerons). The approach of \cite{dip2,loop1}
allows only for pomeron splittings at positive rapidities (above the
dashed line) and only for pomeron mergers at negative rapidities
(below the dashed line). Obviously this approximation consistently
resums all pomeron loops stretching over rapidity interval greater
than $Y/2$, where $Y$ is the total rapidity interval. However, some of
the ``smaller'' pomeron loops, namely the loops stretching over
rapidity intervals less than $Y/2$, are not included consistently. An
example of pomeron loops which are no included is shown in
\fig{ploop1}B, where the pomeron ladder both splits and merges at
positive and at negative rapidities. One may argue that small pomeron
loops are always subleading at high energies and can be safely
neglected compared to larger loops. Still, the small loops of
\fig{ploop1}B become order $1$ corrections at rapidity defined by
\be\label{2yu}
\as^2 \, e^{(\alpha_P - 1) \, {\tilde Y}_U /2} \, \sim \, 1,
\ee
which gives
\be
{\tilde Y}_U \, = \, 2 \, Y_U.  
\ee
Therefore, we conclude that large pomeron loops of \cite{dip2,loop1}
are a good approximation to high energy onium-onium scattering cross
section only in the rapidity interval
\be\label{limits}
Y_U \, \le \, Y \, < \, 2 \, Y_U.
\ee

But what happens at higher rapidities, for $Y \, \ge \, 2 \, Y_U$?
Would resummation of smaller pomeron loops of \fig{ploop1}B be
sufficient to consistently calculate the total onium-onium scattering
cross section? Our answer is ``no''. The reason for this conclusion is
that other (non--pomeron loop) higher order corrections also become
important at $Y \, \ge \, 2 \, Y_U$. For instance, NLO corrections to
pomeron impact factor lead to a contribution of the order of
\cite{BCGK}
\be\label{hoc1}
\as^3 \, e^{(\alpha_P - 1) \, Y}
\ee
coming from the whole ladder. One may also speculate that inserting
one rung of NLO BFKL kernel \cite{NLOBFKL} into LO BFKL ladder
\cite{KM2} would generate the contribution of the same order as shown
in \eq{hoc1}. The corrections from \eq{hoc1} would become of order $1$
at rapidity $(3/2) \, Y_U$. This would lead to an even smaller upper
bound in \eq{limits}, which would become $(3/2) \, Y_U$ instead of $2
\, Y_U$. The resulting applicability window for large
pomeron loops is then limited to
\be\label{limits1}
Y_U \, \le \, Y \, < \, \frac{3}{2} \, Y_U.
\ee 
Indeed, since $Y_U \gg 1$ when $\as \ll 1$, the rapidity window in
\eq{limits1} would still be very large for asymptotically small
coupling. Note that, for the values of the strong coupling constant
accessible in the current accelerators, NLO BFKL corrections are
numerically large and would modify our above conclusions.

At rapidity $Y= 2 \, Y_U$ other corrections will become important as
well. In performing the integration over rapidities of a leading order
pomeron loop the lower limit of integration gives a loop stretching
over zero rapidity interval \cite{NP,yuri}. Such diagram would still
have a single pomeron exchange, giving $e^{(\alpha_P - 1)
\, Y}$, and the zero size loop, giving $\as^4$ due to two pair of
dipoles interacting, overall contributing a factor of
\be\label{hoc2}
\as^4 \, e^{(\alpha_P - 1) \, Y}.
\ee
A correction like the one shown in \eq{hoc2} may also result from
inserting NLO BFKL kernel \cite{NLOBFKL} twice in the LO BFKL ladder
\cite{KM2}, or from inserting NNLO kernel once in the LO BFKL ladder.

The corrections in \eq{hoc2} become of order $1$ when $Y = 2 Y_U$. In
other words, these terms become comparable to ``small'' pomeron loops
stretching over $Y_U$ rapidity interval. This is natural since the
resummation parameter for these higher order corrections is easily
expressed in terms of small pomeron loops resummation parameter
\be
\as^4 \, e^{(\alpha_P - 1) \, Y} \, = \, \left[ \as^2 \, 
e^{(\alpha_P - 1) \, Y/2} \right]^2,
\ee
so that both become comparable to $1$ simultaneously. 

We can conclude that pomeron loop approximation is only valid in the
rapidity region specified by \eq{limits1}. To go beyond that rapidity
interval one has to include a scope of other subleading non-pomeron
loop effects listed above. The corrections of \eq{hoc2} were recently
discussed in \cite{loop6} to account for the apparent discrepancy
between the results of \cite{loop2} and \cite{loop5}, indicating that
such corrections can not be consistently resummed within the current
pomeron loop resummation formalisms.

\begin{figure}
\begin{center}
\epsfxsize=17cm
\leavevmode
\hbox{\epsffile{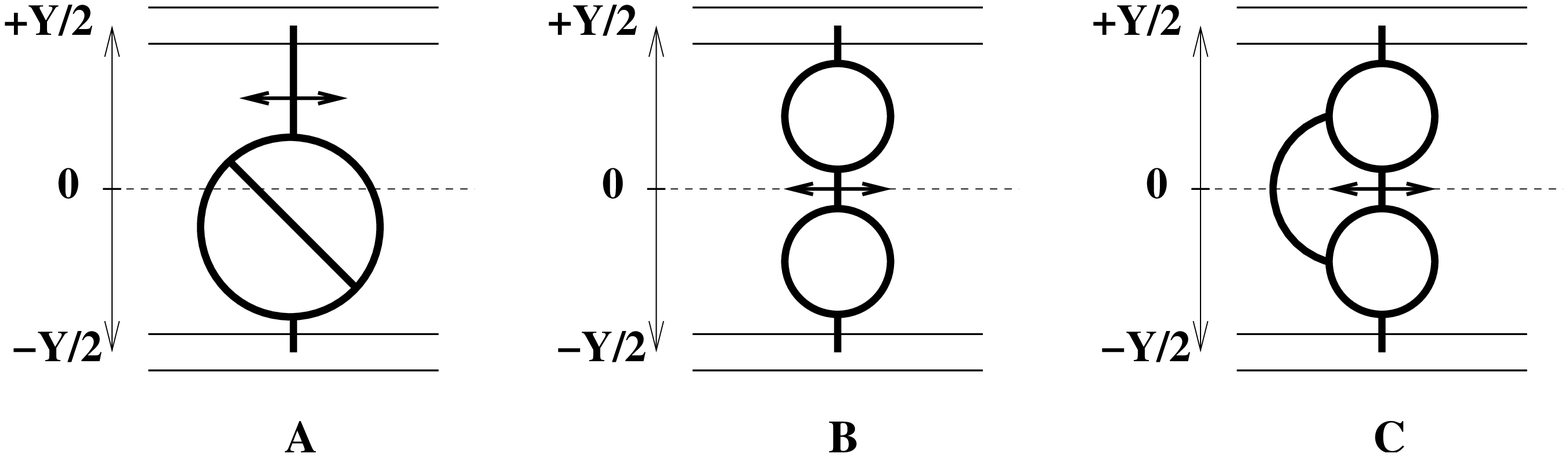}}
\end{center}
\caption{Gluon production diagrams for onium-onium scattering: (A) a 
typical diagram contributing to \eq{incl} above; (B) contributions of
smaller pomeron loops still giving IR-singular gluon production cross
section; (C) contributions of smaller pomeron loops which may lead to
IR-regular gluon production cross section. Again thick lines denote
BFKL ladder and outgoing arrows denote produced gluons. }
\label{ploop2}
\end{figure}

Now let us return to the question of gluon production in onium-onium
scattering. A typical diagram contributing to the above result
(\ref{incl}) is shown in \fig{ploop2}A. There, the produced gluons are
denoted by outgoing double arrows and the BFKL ladders are again
denoted by thick lines. The graph in \fig{ploop2}A has linear BFKL
evolution (single ladder) on one side of the produced gluon (in
rapidity), with all the nonlinear pomeron loop effects being on the
other side of the produced gluon, just like we have obtained in
deriving \eq{incl}. The resulting cross section is IR divergent as
$\sim 1/k_T^2$ and is valid in the rapidity window given by
\eq{limits1}. Therefore, the $\sim 1/k_T^2$ divergence persists as long
as pomeron loop approximation is valid.

Small pomeron loop corrections of \fig{ploop1}B become important at
rapidities $Y \ge 2 \, Y_U$. They may lead to gluon production
diagrams like the one shown in \fig{ploop2}B. Such graphs include
non-linear saturation effects in the wave functions of both
onia. However, they do not violate $k_T$-factorization, as could be
seen in the graph of \fig{ploop2}B, which has no ``cross-talk''
between the onia wave functions other then the interaction leading to
gluon production. Therefore, diagrams like the one shown in
\fig{ploop2}B would still lead to $\sim 1/k_T^2$ divergence in the 
infrared. 

However, one may also imagine small pomeron loop corrections which
{\sl would} violate $k_T$-factorization, as shown in \fig{ploop2}C. As
we discussed in Sect. 3, in the large pomeron loop approximation of
\cite{dip2,loop1} such corrections would cancel due to real-virtual
cancellations of \cite{CM}, making the final result for the inclusive
cross section adhere to AGK cutting rules. Once the small pomeron
loops of \fig{ploop2}B are included in both of the onia wave
functions, such real-virtual cancellations may stop being valid. This
would lead to diagrams in \fig{ploop2}C contributing to gluon
production cross section, and, possibly, regulating the $\sim 1/k_T^2$
divergence. However, such regularization of the divergence would be
due to small pomeron loops, which become important at rapidities $Y
\ge 2 \, Y_U$, when the pomeron loop approximation breaks down. Thus
it may happen that regularization of $\sim 1/k_T^2$ divergence would
have to also include effects going beyond the pomeron loop
approximation, alongside with the purely pomeron loop diagrams of
\fig{ploop2}C.

If, instead of going to higher energies/rapidities, we fix the
rapidity $Y$ and decrease $k_T$, ``higher twist'' effects would become
progressively more important. Our conclusion above implies that, while
higher twist terms due to large pomeron loops of \cite{dip2,loop1} do
not regularize the $\sim 1/k_T^2$ divergence of the inclusive gluon
production cross section, the higher twist terms associated with small
pomeron loops, not included in the approximation of \cite{dip2,loop1},
which may regulate the $\sim 1/k_T^2$ singularity, come in at the same
time as the higher twist terms associated with other (non-pomeron
loop) kinds of corrections.


\section*{Acknowledgments} 

This work is supported in part by the U.S. Department of Energy under
Grant No. DE-FG02-05ER41377.


\renewcommand{\theequation}{A\arabic{equation}}
  \setcounter{equation}{0} 
  \section*{Appendix A}

Starting with the first term in the square brackets of \eq{incl}
(integrated over ${\un B}$) we note that $z_2$ integral can be carried
out using
\be
\int d^2 z \, e^{i {\un k} \cdot {\un z}} \, \frac{{\un z}}{{\un z}^2} 
\, = \, 2 \, \pi \, i \, \frac{{\un k}}{{\un k}^2}
\ee
yielding
\ben
\int d^2 x_{2} \, d^2 x_3 \ n_1 \left( {\un x}_{01}, 
\frac{Y}{2} -y; {\un x}_{23} \right) \, \frac{\as C_F}{2 \, \pi^3} \, i \, 
\int \, d^2 z_1 \, e^{- i {\un k} \cdot ({\un z}_1 - {\un x}_j)} \, 
\sum_{i,j=2}^3 (-1)^{i+j} \frac{{\underline z}_1- {\underline x}_i}{|{\underline z}_1-
{\underline x}_i|^2} \cdot \frac{{\un k}}{{\un k}^2} 
\een
\be\label{A2}
\times \, D_{Gq} \left({\underline z}_1 - {\underline x}_j, {\un x}_{0'1'},
\frac{1}{2} ({\underline z}_1 + {\underline x}_j), y + \frac{Y}{2} \right).
\ee
Defining ${\un z} = {\underline z}_1 - {\underline x}_j$ and rewriting
$d^2 x_{2} \, d^2 x_3 \, = \, d^2 x_{23} \, d^2 b$, where $\un b =
{\un x}_j + (1/2) {\un z}$ we can transform \eq{A2} into
\ben
\int d^2 x_{23} \ n_1 \left( {\un x}_{01}, 
\frac{Y}{2} -y; {\un x}_{23} \right) \, \frac{\as C_F}{2 \, \pi^3} \, i \, 
\int \, d^2 z \, e^{- i {\un k} \cdot {\un z}} \, 
\sum_{i,j=2}^3 (-1)^{i+j} \, \frac{{\underline z} - {\underline x}_{ij}}
{|{\underline z} - {\underline x}_{ij}|^2} \cdot \frac{{\un k}}{{\un k}^2}
\een
\be\label{A3}
\times \, \int d^2 b \,  D_{Gq} \left({\underline z}, {\un x}_{0'1'},
{\underline b}, y + \frac{Y}{2} \right).
\ee
The second term in the square brackets of \eq{incl} gives an identical
contribution.

The third term in the square brackets of \eq{incl} is
different. Defining ${\un {\tilde z}}_1 = {\un z}_1 - {\un x}_i$ and 
${\un {\tilde z}}_2 = {\un z}_2 - {\un x}_j$ we write it as
\ben
- \int d^2 x_{2} \, d^2 x_3 \ n_1 \left( {\un x}_{01}, 
\frac{Y}{2} -y; {\un x}_{23} \right) \, \frac{\as C_F}{(2 \, \pi)^2 \, \pi^2} \,
\int \, d^2 {\tilde z}_1 \, d^2 {\tilde z}_2 \, 
\frac{{\underline {\tilde z}}_1}{{\underline {\tilde z}}_1^2} \cdot
\frac{{\underline {\tilde z}}_2}{{\underline {\tilde z}}_2^2} \, 
\sum_{i,j=2}^3 (-1)^{i+j} \, e^{- i {\un k} \cdot ({\un {\tilde z}}_1 - 
{\un {\tilde z}}_2 + {\un x}_{ij})}  
\een
\be\label{A4}
\times \, D_{Gq} \left({\underline {\tilde z}}_1 - {\underline {\tilde z}}_2 
+ {\un x}_{ij}, {\un x}_{0'1'},
\frac{1}{2} ({\underline {\tilde z}}_1 + {\underline {\tilde z}}_2) + 
\frac{1}{2} ({\un x}_i + {\un x}_j), y + \frac{Y}{2} \right).
\ee
Dropping tildes over ${\un z}_1$ and ${\un z}_2$, rewriting $d^2 x_{2}
\, d^2 x_3 \, = \, d^2 x_{23} \, d^2 b$ with
\ben
{\un b} \, = \, \frac{1}{2} ({\un x}_i + {\un x}_j) + 
\frac{1}{2} ({\underline {\tilde z}}_1 + {\underline {\tilde z}}_2)
\een
we recast \eq{A4} into
\ben
- \int d^2 x_{23} \ n_1 \left( {\un x}_{01}, 
\frac{Y}{2} -y; {\un x}_{23} \right) \, \frac{\as C_F}{(2 \, \pi)^2 \, \pi^2} \,
\int \, d^2 z_1 \, d^2 z_2 \, 
\frac{{\underline z}_1}{{\underline z}_1^2} \cdot
\frac{{\underline z}_2}{{\underline z}_2^2} \, 
\sum_{i,j=2}^3 (-1)^{i+j} \, e^{- i {\un k} \cdot ({\un z}_1 - 
{\un z}_2 + {\un x}_{ij})}  
\een
\be\label{A5}
\times \, \int d^2 b \, 
D_{Gq} \left({\underline z}_1 - {\underline z}_2 
+ {\un x}_{ij}, {\un x}_{0'1'}, {\un b}, y + \frac{Y}{2} \right).
\ee
Defining ${\un z} = {\un z}_1 - {\un z}_2 + {\un x}_{ij}$ and
integrating over ${\un y} = (1/2) ({\un z}_1 + {\un z}_2)$ using
\be
\int d^2 y \, \frac{{\underline y}}{{\underline y}^2} \cdot
\frac{{\un y} + {\underline z}}{|{\un y} + {\underline z}|^2} \, = \, 
2 \, \pi \, \ln \frac{1}{|{\un z}| \, \mu}
\ee
with $\mu$ some infrared cutoff yields
\ben
- \int d^2 x_{23} \ n_1 \left( {\un x}_{01}, 
\frac{Y}{2} -y; {\un x}_{23} \right) \, \frac{\as C_F}{(2 \, \pi) \, \pi^2} \,
\int \, d^2 z \, 
\sum_{i,j=2}^3 (-1)^{i+j} \, e^{- i {\un k} \cdot {\un z}}  \, 
\ln \left( \frac{1}{|{\un z} - {\un x}_{ij}| \, \mu} \right)
\een
\be\label{A7}
\times \, \int d^2 b \, 
D_{Gq} \left({\underline z}, {\un x}_{0'1'}, {\un b}, y + \frac{Y}{2}
\right).
\ee

Finally, similar to the above, the last term in the square brackets of
\eq{incl} can be written as
\ben
- \int d^2 x_{23} \ n_1 \left( {\un x}_{01}, 
\frac{Y}{2} -y; {\un x}_{23} \right) \, \frac{\as C_F}{\pi^2} \, 
\frac{1}{{\un k}^2} \, \int \, d^2 z \, 
\sum_{i,j=2}^3 (-1)^{i+j} \, e^{- i {\un k} \cdot {\un z}}  \, 
\delta^2 ({\un z} - {\un x}_{ij})
\een
\be\label{A8}
\times \, \int d^2 b \, 
D_{Gq} \left({\underline z}, {\un x}_{0'1'}, {\un b}, y + \frac{Y}{2}
\right).
\ee

Combining Eqs. (\ref{A3}) (twice), (\ref{A7}) and (\ref{A8}) yields
\ben
\frac{d \sigma ({\un x}_{01}, {\un x}_{0'1'})}{d^2 k \, dy} 
\, = \, \int d^2 x_{23} \ n_1 \left( {\un x}_{01}, 
\frac{Y}{2} -y; {\un x}_{23} \right) \, \frac{\as C_F}{2 \, \pi^3} \, 
\frac{1}{{\un k}^2} \, \int \, d^2 z \, 
\sum_{i,j=2}^3 (-1)^{i+j} \, e^{- i {\un k} \cdot {\un z}}
\een
\be\label{A9}
\times \, \int d^2 b \, 
D_{Gq} \left({\underline z}, {\un x}_{0'1'}, {\un b}, y + \frac{Y}{2}
\right) \, \left[2 \, i \, {\un k} \cdot \, 
\frac{{\underline z} - {\underline x}_{ij}}
{|{\underline z} - {\underline x}_{ij}|^2} - {\un k}^2 \, \ln \left(
\frac{1}{|{\un z} - {\un x}_{ij}| \, \mu} \right) - 2 \, \pi \,
\delta^2 ({\un z} - {\un x}_{ij}) \right],
\ee
which can be rewritten as
\ben
\frac{d \sigma ({\un x}_{01}, {\un x}_{0'1'})}{d^2 k \, dy} 
\, = \, \int d^2 x_{23} \ n_1 \left( {\un x}_{01}, 
\frac{Y}{2} -y; {\un x}_{23} \right) \, \frac{\as C_F}{2 \, \pi^3} \, 
\frac{1}{{\un k}^2} \, \int \, d^2 z \, d^2 b \, 
D_{Gq} \left({\underline z}, {\un x}_{0'1'}, {\un b}, y + \frac{Y}{2}
\right)
\een
\be\label{A10}
\times \,  \sum_{i,j=2}^3 (-1)^{i+j} \, \nabla_z^2 
\left[  e^{- i {\un k} \cdot {\un z}} \, \ln \left( \frac{1}{|{\un
z} - {\un x}_{ij}| \, \mu} \right) \right],
\ee
where $\nabla_z^2$ is the transverse coordinate (2d) gradient
squared. Integrating by parts yields
\ben
\frac{d \sigma ({\un x}_{01}, {\un x}_{0'1'})}{d^2 k \, dy} 
\, = \,  \frac{\as C_F}{2 \, \pi^3} \, 
\frac{1}{{\un k}^2} \, \int \, d^2 z \,  \left[ \nabla_z^2 \, \int d^2 b \
D_{Gq} \left({\underline z}, {\un x}_{0'1'}, {\un b}, y + \frac{Y}{2}
\right) \right] \, e^{- i {\un k} \cdot {\un z}}
\een
\be\label{A11}
\times \,  \sum_{i,j=2}^3 (-1)^{i+j} \, \int d^2 x_{23} \ n_1 \left( {\un x}_{01}, 
\frac{Y}{2} -y; {\un x}_{23} \right) \, \ln \left( \frac{1}{|{\un
z} - {\un x}_{ij}| \, \mu} \right)
\ee
and, remembering that
\be 
\nabla_z^2 \left( {\un z}^2 \, \ln  \frac{1}{|{\un z}| \, \mu} \right) \, = \, 
4 \, \ln  \frac{1}{|{\un z}| \, \mu}
\ee
reduces \eq{A11} to \eq{s1} as desired.

\end{document}